\documentclass{article}

\usepackage{PRIMEarxiv}
\usepackage[utf8]{inputenc} 
\usepackage[T1]{fontenc}    
\usepackage{booktabs}       
\usepackage{amsfonts}       
\usepackage{amsmath}
\usepackage{bbding}
\usepackage{nicefrac}       
\usepackage{microtype}      
\usepackage{lipsum}
\usepackage{color,xcolor}
\usepackage{subfiles} %
\usepackage{makecell} %
\usepackage{multirow} %
\usepackage{colortbl}
\usepackage{pifont}%
\usepackage{algorithm}      
\usepackage{algorithmic}    %

\usepackage[misc]{ifsym}
\usepackage{graphicx}       
\usepackage{subfigure}
\usepackage{multirow}
\usepackage{overpic}
\usepackage{enumitem}
\usepackage{caption}
\usepackage{subcaption}
\graphicspath{{figures/}}     
\definecolor{shapecolor}{rgb}{0.1,0.45,0.8}

\definecolor{arylideyellow}{rgb}{0.91, 0.84, 0.42}

\definecolor{lblue}{rgb}{0.1,0.45,0.8}
\definecolor{lgreen}{rgb}{0.18,0.58,0.18}

\newcommand{\figref}[1]{Figure.~\ref{#1}}
\newcommand{\tabref}[1]{Table.~\ref{#1}}

\definecolor{cvprblue}{rgb}{0.21,0.49,0.74}
\usepackage[pagebackref,breaklinks,colorlinks,citecolor=cvprblue]{hyperref}

\newenvironment{myitemize}[1][]{
\begin{list}{{#1}} 
    {
     \setlength{\leftmargin}{0cm}      %
     \setlength{\parsep}{0ex}          %
     \setlength{\topsep}{0bp}          %
     \setlength{\itemsep}{0ex}         %
     \setlength{\labelsep}{0.2em}      %
     \setlength{\itemindent}{2em}      %
     \setlength{\listparindent}{1.5em} %
    }}
{\end{list}}

\title{MHS-VM: Multi-Head Scanning in Parallel Subspaces for Vision Mamba}

\author{
  Zhongping Ji \\
   \\
  jzp@hdu.edu.cn \\
  Hangzhou Dianzi University}

\begin{document}
\maketitle

\begin{abstract}
Recently, State Space Models (SSMs), with Mamba as a prime example, have shown great promise for long-range dependency modeling with linear complexity. Then, Vision Mamba and the subsequent architectures are presented successively, and they perform well on visual tasks. The crucial step of applying Mamba to visual tasks is to construct 2D visual features in sequential manners. To effectively organize and construct visual features within the 2D image space through 1D selective scan, we propose a novel Multi-Head Scan (MHS) module. The embeddings extracted from the preceding layer are projected into multiple lower-dimensional subspaces. Subsequently, within each subspace, the selective scan is performed along distinct scan routes. The resulting sub-embeddings, obtained from the multi-head scan process, are then integrated and ultimately projected back into the high-dimensional space. Moreover, we incorporate a Scan Route Attention (SRA) mechanism to enhance the module's capability to discern complex structures. To validate the efficacy of our module, we exclusively substitute the 2D-Selective-Scan (SS2D) block in VM-UNet with our proposed module, and we train our models from scratch without using any pre-trained weights. The results indicate a significant improvement in performance while reducing the parameters of the original VM-UNet. The code for this study is publicly available at \textit{\textcolor{lblue}{https://github.com/PixDeep/MHS-VM}}.
\end{abstract}


\section{Introduction}

In recent years, the relentless evolution of deep learning has propelled substantial advancements in the field of computer vision. Visual representation learning stands as a pivotal step in computer vision. Previously, two predominant classes of foundational models, namely Convolutional Neural Networks (CNNs) \cite{unet2015, Resnet2016, densenet2017,  liu2022convnet} and Vision Transformers (ViTs) \cite{vit2021, Swin2021, DeiT2021}, have been widely utilized across a spectrum of visual tasks. Both models have garnered notable achievements in generating expressive visual representations, yet ViTs often surpass CNNs in performance, a feat that can be ascribed to their global receptive fields and the dynamic weighting enabled by the attention mechanism. However, the complexity of attention mechanisms grows quadratically with image size, imposing a significant computational load on dense prediction tasks like semantic segmentation and object detection. Therefore, it is imperative to design visual models with linear complexity, while still preserving the advantages associated with global receptive fields and dynamic weights. 

Recently, the field of computer vision has seen a resurgence of interest in State Space Models (SSMs) \cite{mambagu2023, gssmehta2023, wang2023}, which have traditionally excelled in modeling long-range dependencies with their linear computational scaling. Advancements in SSMs extend their application to complex visual tasks where they promise to offer efficiency gains over established architectures like CNNs and ViTs. The Mamba model \cite{mambagu2023}, leveraging data-dependent SSM layers, demonstrated its prowess by surpassing Transformers across various scales on large-scale real-world datasets, maintaining linear computational complexity in sequence length. This achievement underscores Mamba's potential as a transformative architecture for language modeling, suggesting that SSMs can be competitive alternatives to Transformers in domains beyond their original application. Inspired by Mamba's success, Vision Mamba (Vim) \cite{zhu2024vision} was introduced, marking a significant step towards applying pure SSM-based methods to visual tasks without reliance on attention mechanisms. Vim, by integrating bidirectional selective state space models and position embeddings, enhances visual representation learning, achieving superior performance to DeiT \cite{DeiT2021} on ImageNet classification and dense prediction tasks such as object detection and semantic segmentation. Meanwhile, VMamba \cite{liu2024vmamba} extends the SSM paradigm to visual tasks, focusing on enhancing efficiency and scalability in visual representation learning. VMamba incorporates a Cross-Scan Module (CSM) to enable 1D selective scanning in 2D image space, maintaining a global receptive field while reducing computational complexity from quadratic to linear. This design, along with architectural refinements, makes VMamba as a competitive backbone model that matches or surpasses popular vision models such as ResNet \cite{Resnet2016}, ViT \cite{vit2021}, and Swin \cite{Swin2021} in performance. Furthermore, U-Mamba \cite{Umamba2024} addresses the need for efficient long-range dependency modeling in biomedical image segmentation. By integrating the strengths of CNNs for local feature extraction with SSMs for capturing long-range context, U-Mamba showcases the potential of hybrid architectures in improving segmentation performance across diverse biomedical imaging tasks. 

These developments collectively indicate that SSM-based models are maturing into versatile and efficient backbones for visual representation learning, rivaling and in some cases outperforming conventional architectures. Their ability to handle long-range dependencies efficiently, while maintaining or improving performance, suggests a promising trajectory for future research and applications in computer vision, particularly as the scale of visual data and the demand for efficient processing continue to grow. Experimental results show that SSM-based models achieve promising performance across various visual tasks including image classification \cite{chen2024rsmamba,huang2024localmamba}, semantic segmentation \cite{ruan2024vmunet,liao2024lightmunet,wu2024hvmunet,wu2024ultralight,wang2024large}, object detection \cite{huang2024localmamba,wang2024memorymamba}, image restoration \cite{guo2024mambair} and image generation \cite{hu2024zigma,yan2023diffusion,teng2024dim}, etc. 

This paper focuses on exploring the performance of pure SSM-based models on visual tasks. VM-UNet \cite{ruan2024vmunet} is the pioneering model that incorporates VMamba into the U-Net architecture, dedicated to medical image segmentation and leveraging the pure SSM-based approach. Despite some advancements, pure SSM-based model has not yet significantly outperformed traditional methods such as CNNs and ViTs in visual tasks. Particularly when the pure SSM-based model is used exclusively on smaller datasets, it is observed that there is still potential for enhancing its generalization capabilities. This paper suggests optimizing the crucial factor in SSM-based vision models. A crucial step in employing SSMs for various visual tasks involves the transformation of 2D image data into 1D sequences. This conversion is essential for facilitating the processing capabilities of Mamba, which are inherently designed to handle sequential data. By restructuring the image data in this manner, Mamba can effectively analyze and interpret the visual information, leading to improved performance in a multitude of visual recognition tasks. Traditional approaches, like unidirectional scanning in 1D sequences, struggle to simultaneously capture multi-directional dependencies in 2D images, restricting their receptive fields and potentially compromising accuracy in dense prediction tasks such as object detection and segmentation. We attempt to capture multi-directional dependencies in 2D image patches by introducing a Multi-Head Scan (MHS) module. A single scan head is specialized in capturing structural information by adhering to a specific scan pattern within an embedding subspace. In addition, we propose a novel mechanism that prompts the features extracted by the module to implicitly incorporate positional information along the scan route.

In a nutshell, this paper focuses on enhancing the performance of 2D-Selective-Scan (SS2D) in the VSS block of VM-UNet \cite{ruan2024vmunet}. The main contributions of this paper are summarized below:

\begin{myitemize}[\textbullet]

\item[+] A Multi-Head Scan (MHS) mechanism is introduced to enhance visual representation learning. 

\item[+] A richer array of scan patterns is introduced to capture the diverse visual patterns present in vision data.

\item[+] A Scan Route Attention (SRA) mechanism is introduced to enable the model to attenuate or screen out trivial features, thereby enhancing its ability to capture complex structures in images.

\item[+] We develop an easy-to-use module that has exhibited enhanced generalization abilities in segmentation tasks on small-scale medical image datasets.

\end{myitemize}

\section{Preliminaries}
\label{sec:headings}

In modern models based on Structured State Space (SSM), such as Structured State Space Sequence Models (S4) and Mamba, a classical continuous system is utilized. This system maps a one-dimensional input function or sequence, denoted as $x(t) \in \mathcal{R}$, through intermediate implicit states $h(t) \in \mathcal{R}^{N}$ to produce an output $y(t) \in \mathcal{R}$. This process can be encapsulated by a linear Ordinary Differential Equation (ODE):

\begin{equation}
  \label{eqn:ode}
  \begin{aligned}
  h^{\prime}(t) &= \mathbf{A} h(t) + \mathbf{B}x(t) \\
  y(t) &= \mathbf{C} h(t)
  \end{aligned}
\end{equation}

where $\mathbf{A} \in \mathcal{R}^{N\times N}$ is the state matrix, while $\mathbf{B} \in \mathcal{R}^{N \times 1}$ and $\mathbf{C} \in \mathcal{R}^{N\times 1}$ represent the projection parameters.

\textbf{Discretization.} To adapt this continuous system for deep learning applications, S4 and Mamba discretize it by incorporating a timescale parameter $\mathbf{\Delta}$. They transform the continuous parameters $\mathbf{A}$ and $\mathbf{B}$ into discrete counterparts $\overline{\mathbf{A}}$ and $\overline{\mathbf{B}}$ using a fixed discretization rule, such as the zero-order hold (ZOH). This discretization is expressed as:

\begin{equation}
\label{eqn:discretization}
\begin{aligned}
\overline{\mathbf{A}} &= \exp(\mathbf{\Delta} \mathbf{A}) \\
\overline{\mathbf{B}} &= (\mathbf{\Delta} \mathbf{A})^{-1} (\exp(\mathbf{\Delta} \mathbf{A}) - \mathbf{I}) \cdot \mathbf{\Delta} \mathbf{B}
\end{aligned}
\end{equation}

Once discretized, SSM-based models can be computed via either linear recurrence 

\begin{equation}\label{eqn:lr}
\begin{aligned}
h^{\prime}(t) &= \overline{\mathbf{A}} h(t) + \overline{\mathbf{B}} x(t) \\
y(t) &= \mathbf{C} h(t)
\end{aligned}
\end{equation}

or global convolution, 

\begin{equation}\label{eqn:gc}
\begin{aligned}
\overline{\mathbf{K}} &= (\mathbf{C}\overline{\mathbf{B}}, \mathbf{C}\overline{\mathbf{A}} \overline{\mathbf{B}}, \ldots, \mathbf{C}\overline{\mathbf{A}}^{L-1} \overline{\mathbf{B}}) \\
y &= x*\overline{\mathbf{K}}
\end{aligned}
\end{equation}

where $\overline{\mathbf{K}} \in \mathcal{R}^{L}$ denotes a structured convolutional kernel, and $L$ is the length of the input sequence $x$. This method harnesses convolution to simultaneously integrate outputs across the sequence, thereby enhancing computational efficiency and scalability.

\section{Methods}

This section begins with an introduction of the overall architecture of our module. Subsequently, we elaborate on the details of core components.

\begin{figure}[!ht]
\centering
\subfigure{
\begin{minipage}[b]{1.0\linewidth}
\includegraphics[width=\linewidth]{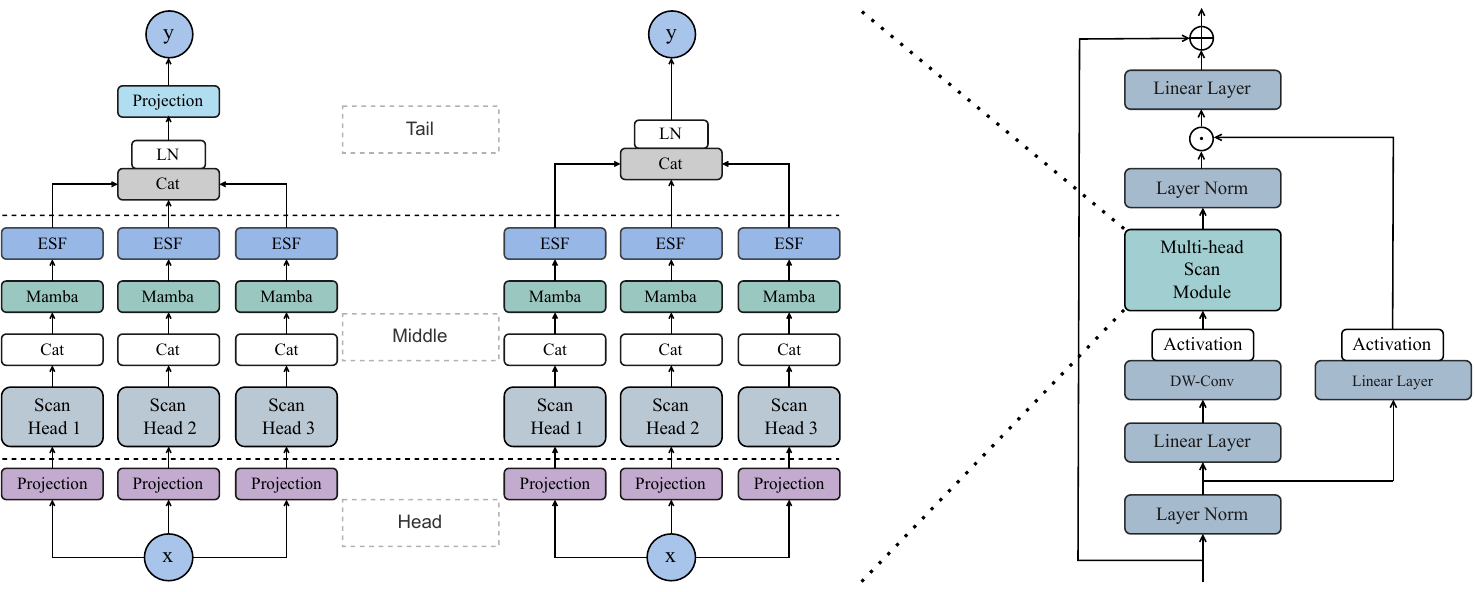} 
\end{minipage}
}
\caption{The architecture of our Multi-Head Scan (MHS) modules. In the illustrated modules, there are three scan headers, and this quantity can be adjusted to suit practical requirements. This design facilitates its application, and we can immediately replace the SS2D module in VSS block of VM-UNet with our MHS module.}
\label{fig:modules}
\end{figure}

The overall architecture of the proposed module is illustrated in \figref{fig:modules}. We have crafted two architectures with subtle differences as shown in the figure. The one on the right, as a variant, eliminates the projection in the tail part, and its performance implications will be discussed in the subsequent ablation study section. Turning our attention to the structure of our module, it is comprised of three principal components: the head, the middle, and the tail, each playing a critical role in the overall architecture. Subsequently, we will offer a comprehensive elucidation of these components.

\subsection{Subspace and Scan Pattern}

The head part focuses on transforming the embedding extracted from the preceding layer into $n$ sub-embeddings which are depicted in $n$ parallel subspaces of equivalent dimension. This can be concisely represented by the following equation:

\begin{equation}
x^1, x^2, \cdots, x^n = [W_1, W_2, \cdots, W_n] x
\end{equation}

where $x \in R^{C_l}$, $x^h \in R^{S}$, and $W_h \in R^{S \times C_l}, h = 1,2,\cdots,n$. By default, the dimension of the space in which embedding $x$ resides is $n$ times the dimension of each subspace, with $C_l = n \times S$, but this is not mandatory.

The middle part consists of $n$ scan heads and Embedding Section Fusion (ESF) sub-modules. Subsequently, we will provide an in-depth explanation of this part.

\begin{figure}[!ht]
\hspace{-0.275in}
\subfigure{
\begin{minipage}[b]{0.288\linewidth}
\begin{flushleft}
\includegraphics[width=\linewidth]{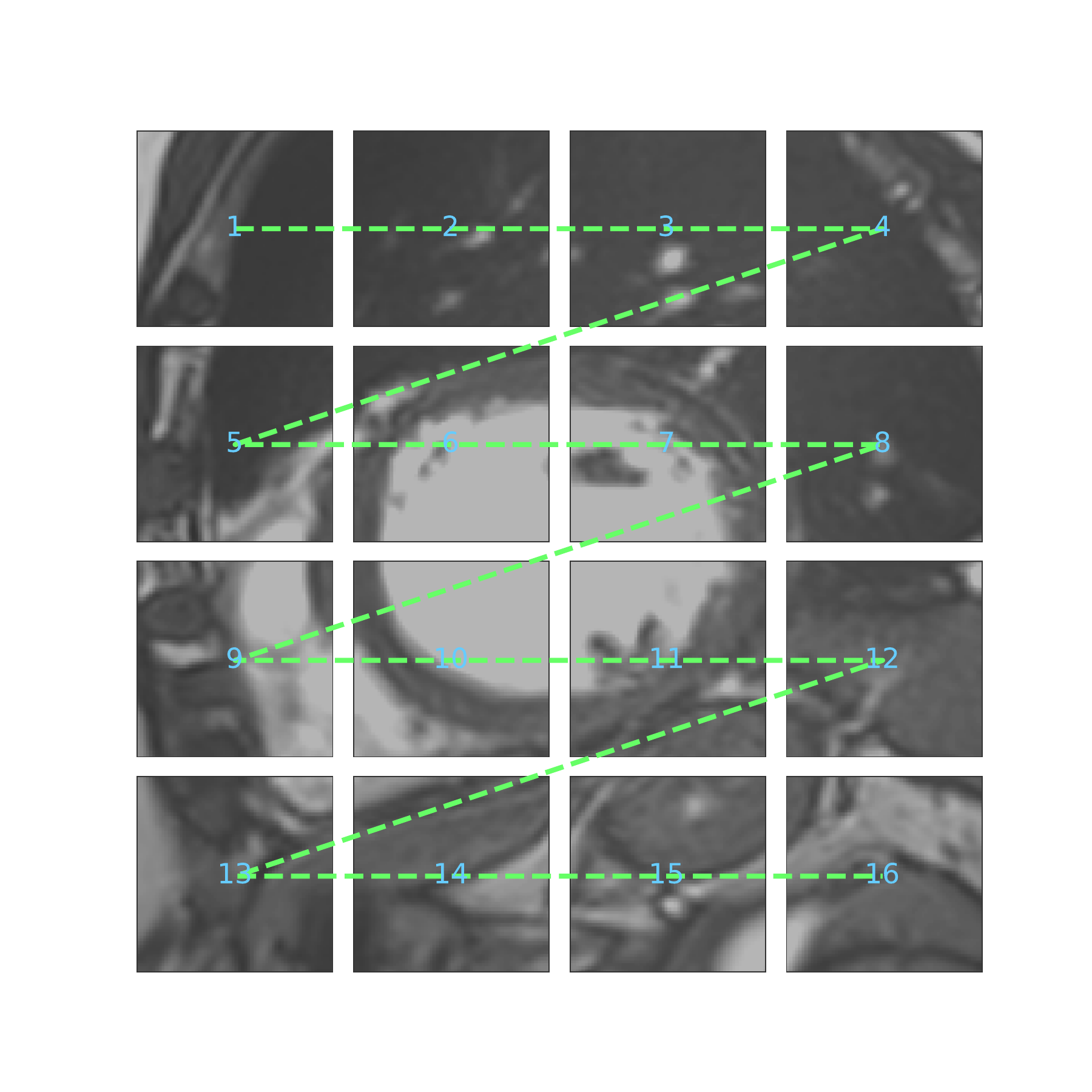}
\end{flushleft}
\end{minipage}
\hspace{-0.258in}
\begin{minipage}[b]{0.288\linewidth}
\includegraphics[width=\linewidth]{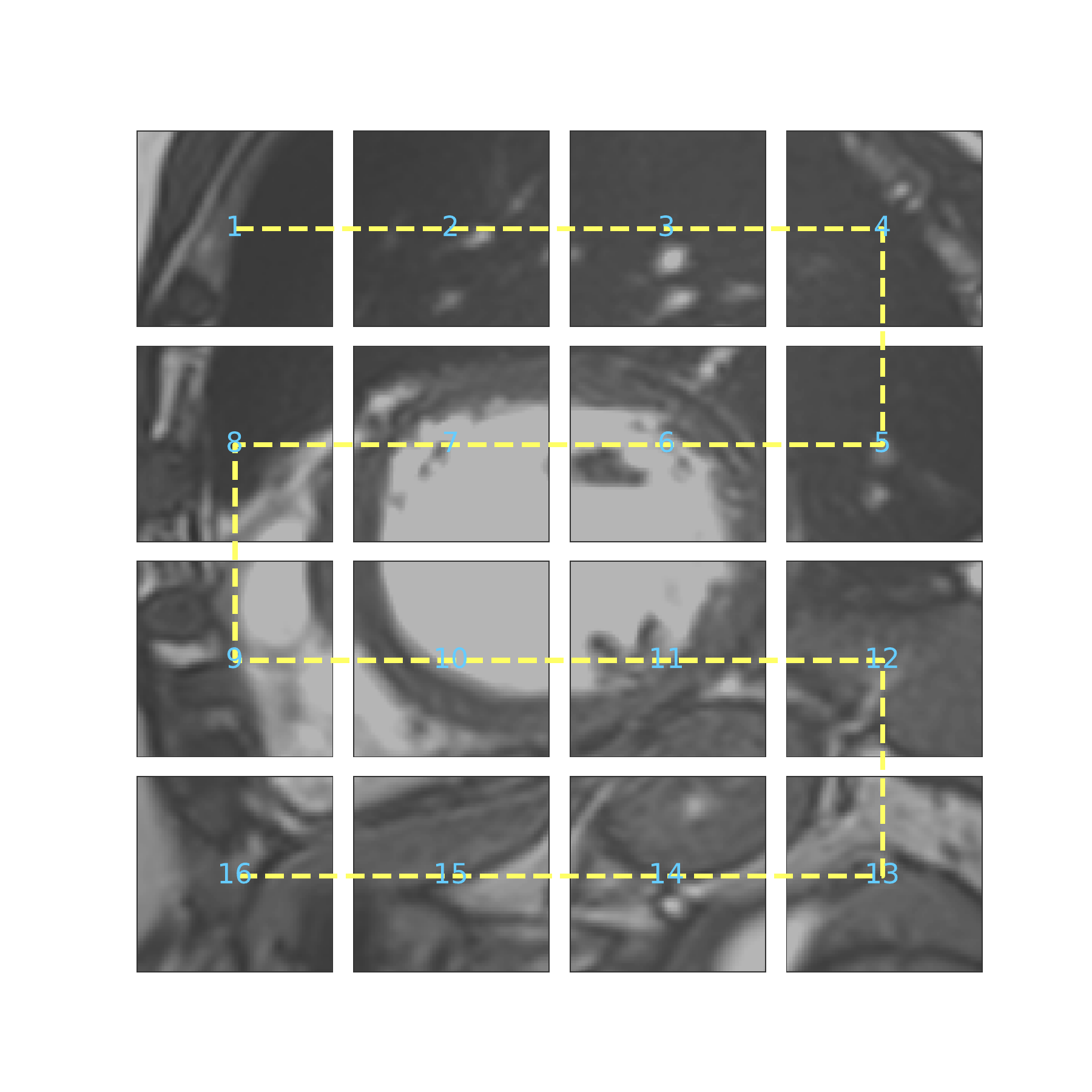}
\end{minipage}
\hspace{-0.258in}
\begin{minipage}[b]{0.288\linewidth}
\includegraphics[width=\linewidth]{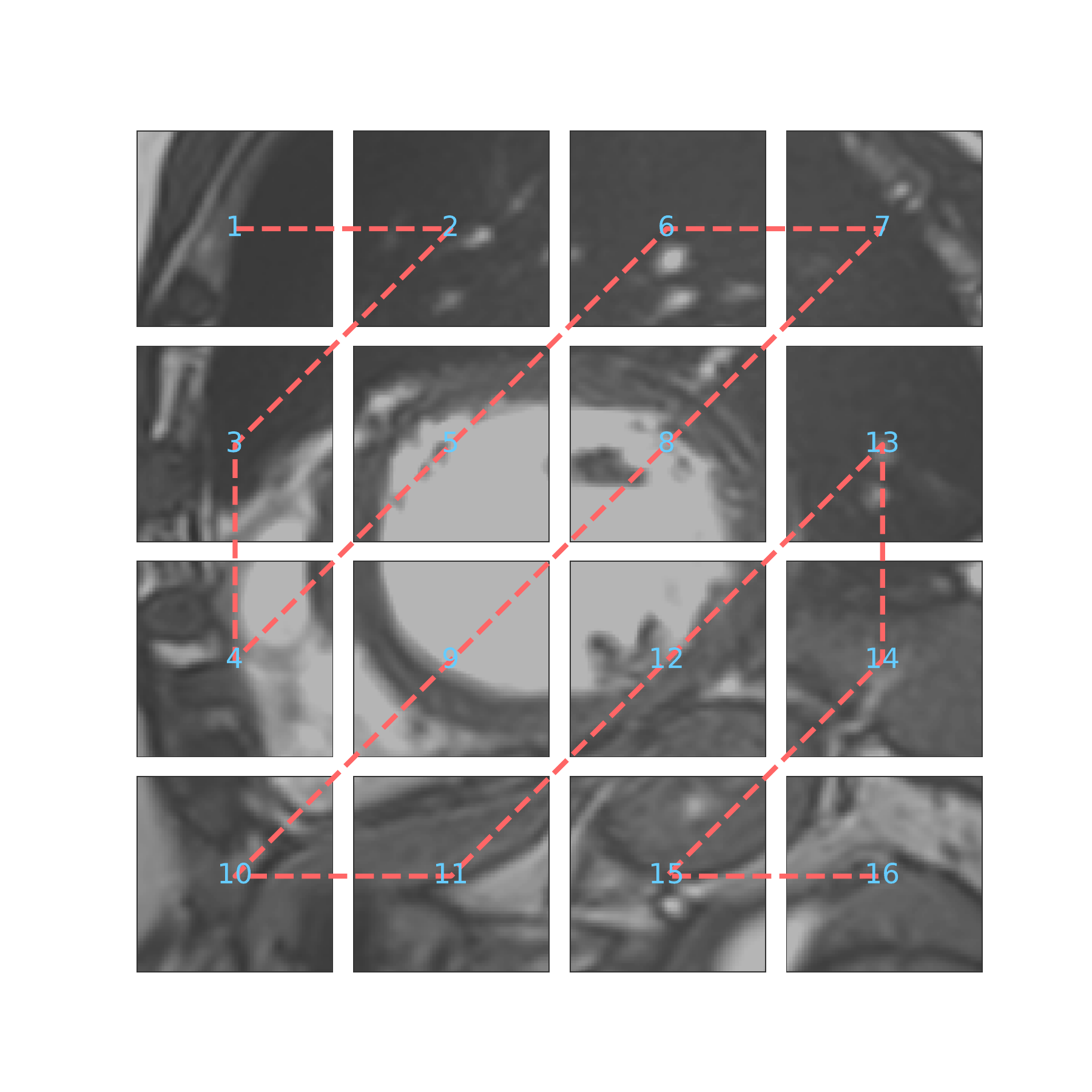}
\end{minipage}
\hspace{-0.258in}
\begin{minipage}[b]{0.288\linewidth}
\includegraphics[width=\linewidth]{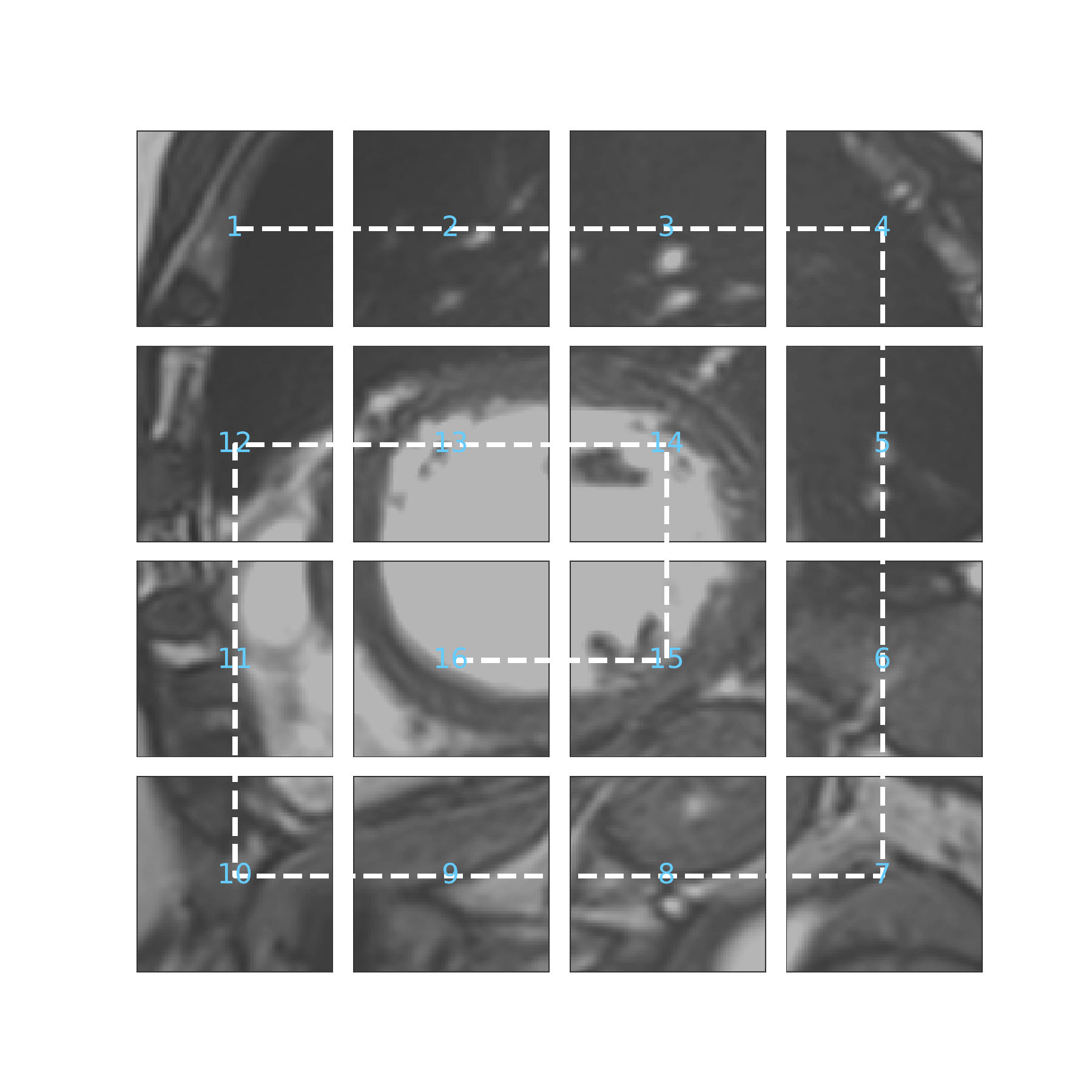}
\end{minipage}
} 
\caption{
Illustration of four scan patterns. From left to right: the traversal paths of image patches for four scan patterns. The number marked along the dotted line indicates the traversal order of the patches along a traversal path.
}
\label{fig:scanpatterns}
\end{figure}

Since the Mamba is designed to process 1D causal data such as language sequences which have causal relationships. However, the intrinsic non-causal nature of 2D vision data presents a substantial challenge to the application of causal processing methods. One idea is to deconstruct 2D vision data through multi-directional 1D sequences. We introduce \textbf{scan pattern} to capture directional dependencies between image patches. As illustrated in \figref{fig:scanpatterns}, this paper explores three additional scan patterns for complex directional dependencies of patches in a 2D image. In addition to the first scan pattern used by the previous work, the other three patterns are more concerned with adjacency between patches. The second pattern is consecutive scanning in horizontal or vertical directions. The third one is consecutive scanning around the diagonals. The last one is consecutive scanning from the outside to the inside spirally. More patterns can be introduced if adjacency between patches is not concerned. In our current architecture, one scan pattern is deployed per subspace. 

\begin{figure}[!ht]
\hspace{-0.05in}
\subfigure{
\begin{minipage}[b]{1.0\linewidth}
\includegraphics[width=\linewidth]{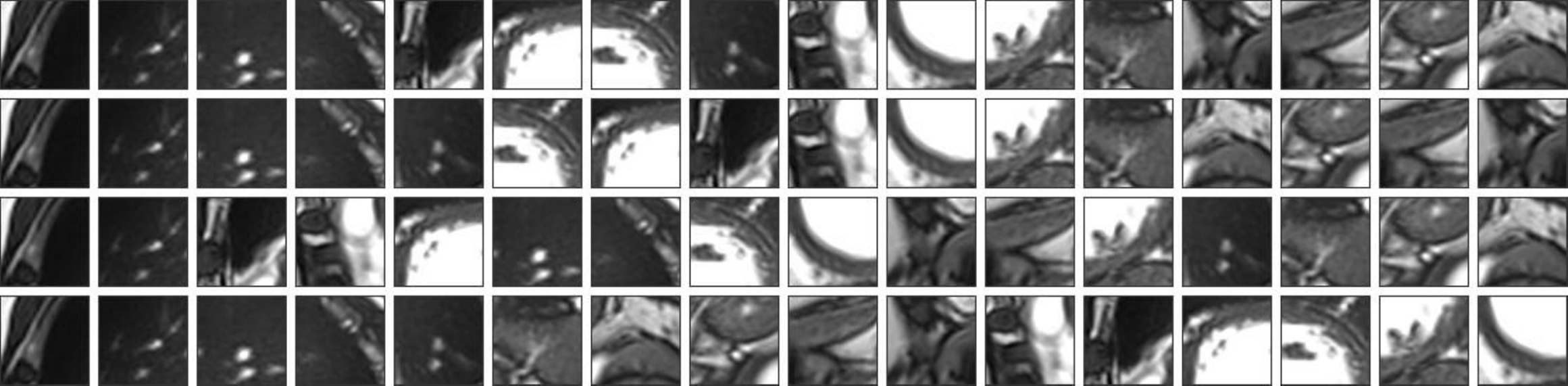}
\end{minipage}
}  
\caption{
Illustration of image patch sequences. Each row displays a 1D sequence of the image patches spread out along a traverse path shown in \figref{fig:scanpatterns}.
}
\label{fig:scanflatten}
\end{figure}

Before input to the Mamba block, we flatten the image (embedding map) into a sequence of patches following the traversal path as shown in \figref{fig:scanflatten}. We can see that each row displays a 1D sequence of image patches spread out along a traverse path. These different 1D causal relationships between patches are combined to approximate the complex structural relationships in 2D vision data.

To facilitate a clearer understanding of our approach, we introduce two key concepts that are central to our patch embedding methodology. First, we denote the consecutive series of patch embeddings extracted along a specific scan route as the \textbf{embedding scan sequence}. This sequence captures the spatial and contextual information of the patches as they are traversed along the scan route. Second, we consider the arrangement of patch embeddings that are flattened along a predefined path, which collectively form a 2D plane, which we term the \textbf{embedding section}. This plane provides a structured representation of the patch embeddings, enabling the model to effectively process and analyze the spatial relationships within the visual data. The green traversal path as shown in \figref{fig:scanpatterns} is used to generate the embedding section for each scan route.

For each scan pattern, we deploy several scan routes to capture diverse directional dependencies between patches. These embedding scan sequences are then fed into a single Mamba block. To independently capture embeddings according to different scan patterns, we refrain from sharing the weights across each Mamba block. This process is expressed by the following formulation:

\begin{equation}
y_{j}^{h} = \operatorname{M}_{h}(x^{h}_{j}),  j = 1,2,\cdots,k; h = 1,2,\cdots,n
\end{equation}

where $\operatorname{M}_{h}$ denotes the $h$-th Mamba block which takes $k$ embedding scan sequences $\{x^{h}_{j} \in R^{C \times L},j=1,2,\cdots,k\}$ gathered from $k$ scan routes as input and produces $k$ output embedding scan sequences $\{y^{h}_{j} \in R^{ C \times L},j=1,2,\cdots,k\}$. $L$ denotes the length of an input embedding scan sequence. These output embedding scan sequences are subsequently rearranged to create embedding sections according to their respective scan routes.

\subsection{Embedding Section Fusion}

\begin{figure}[!ht]
\hspace{-0.275in}
\subfigure{
\begin{minipage}[b]{0.288\linewidth}
\includegraphics[width=\linewidth]{figures/xscan.pdf}
\end{minipage}
\hspace{-0.258in}
\begin{minipage}[b]{0.288\linewidth}
\includegraphics[width=\linewidth]{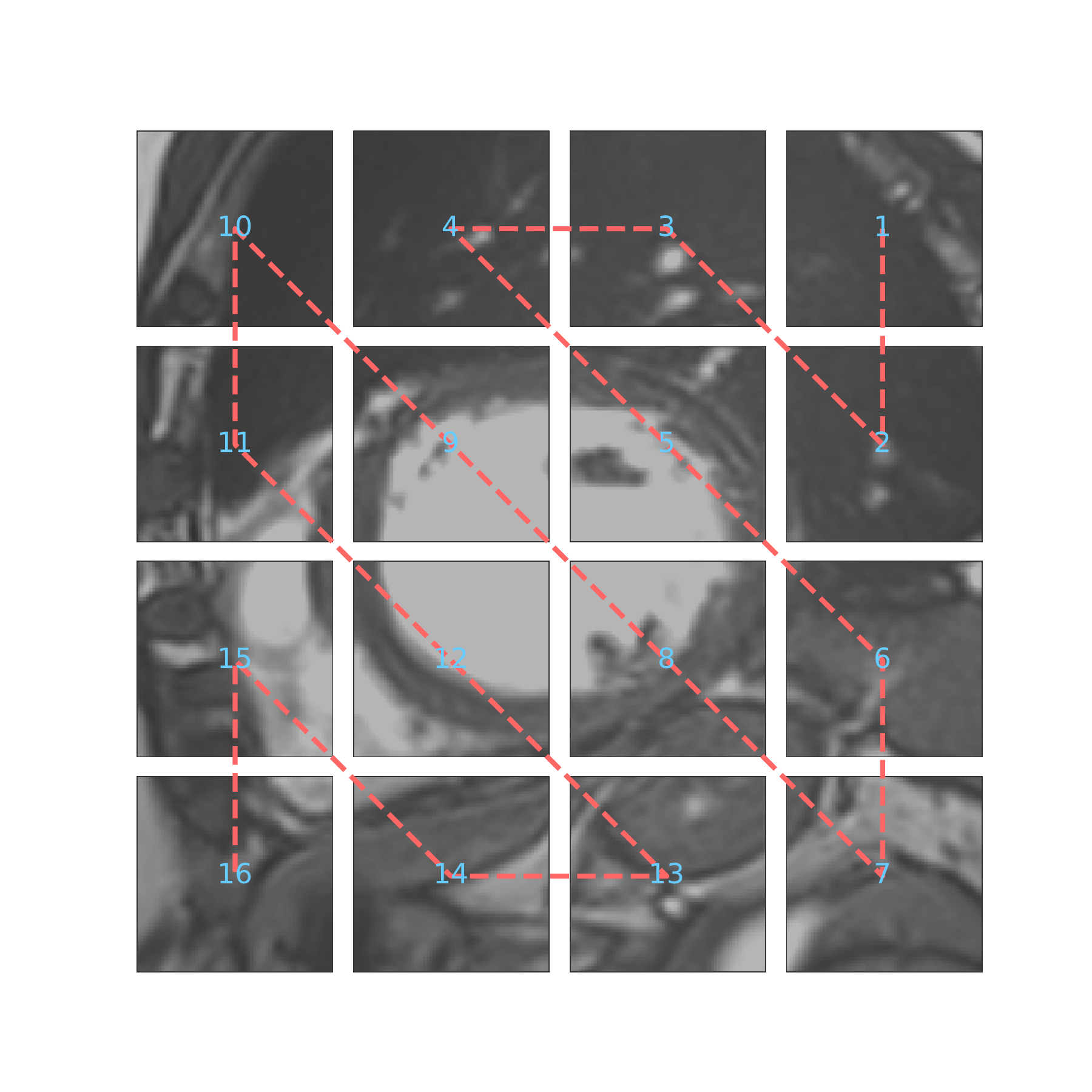}
\end{minipage}
\hspace{-0.258in}
\begin{minipage}[b]{0.288\linewidth}
\includegraphics[width=\linewidth]{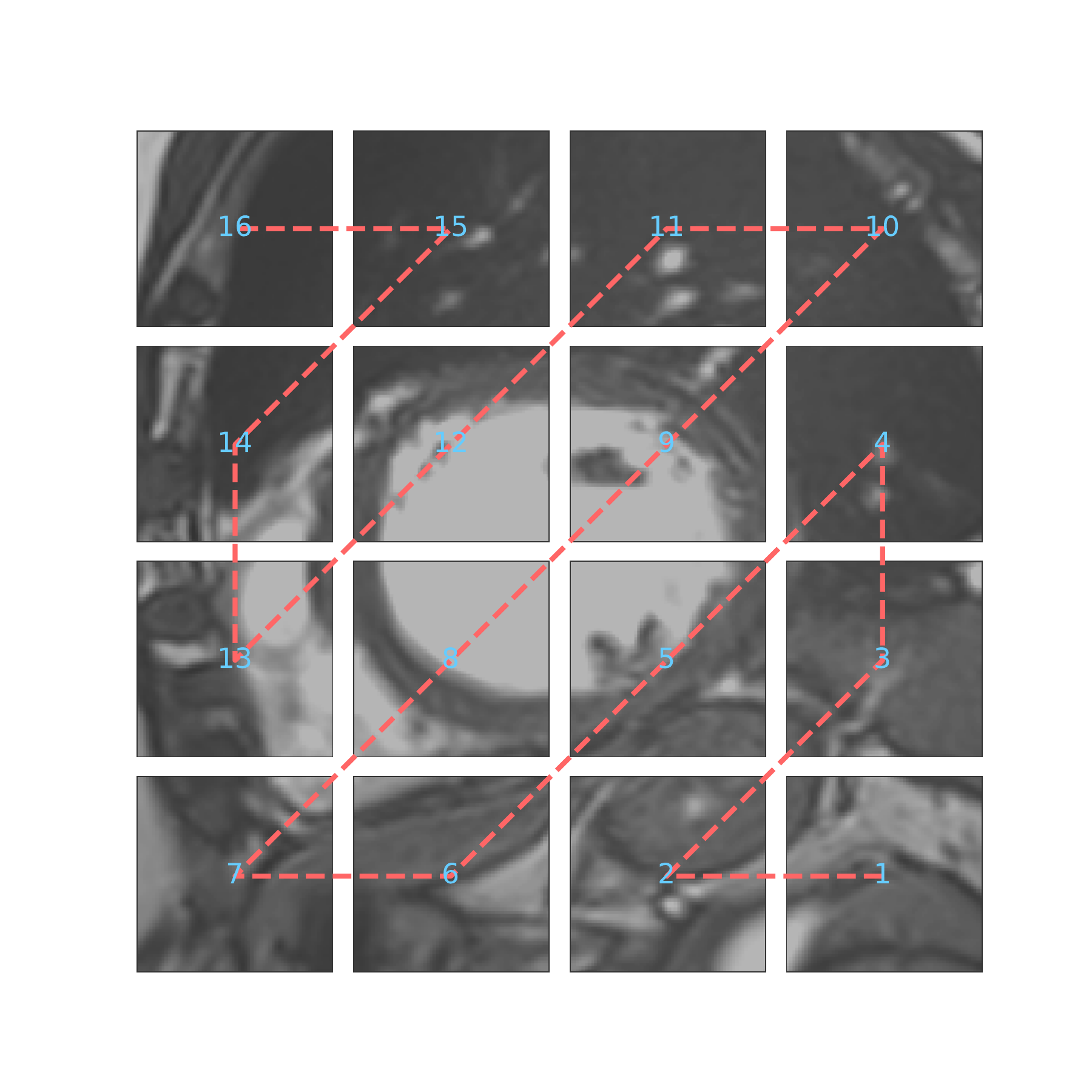}
\end{minipage}
\hspace{-0.258in}
\begin{minipage}[b]{0.288\linewidth}
\includegraphics[width=\linewidth]{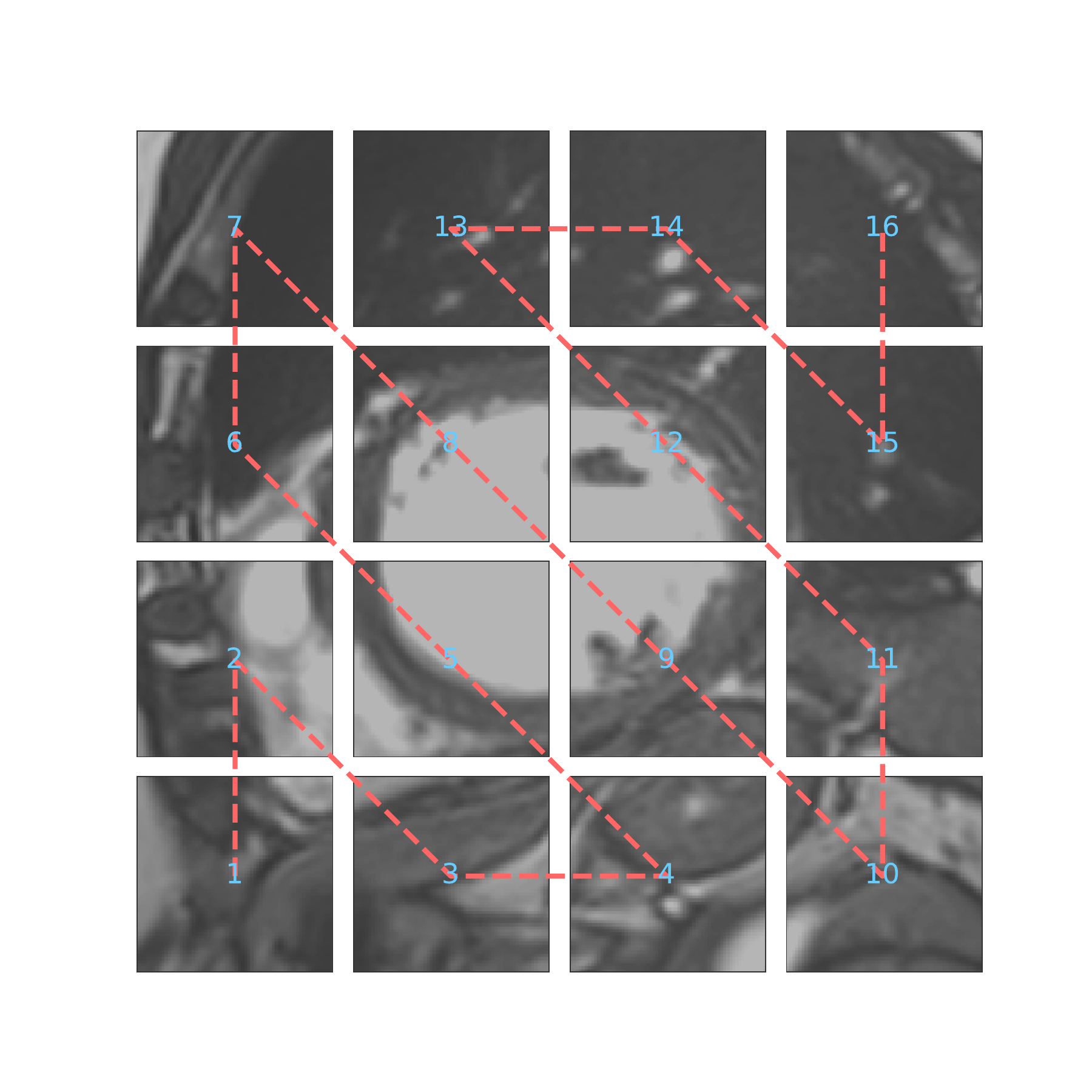}
\end{minipage}
} 
\caption{
Illustration of four scan routes sampled in the third type of scan pattern.
}
\label{fig:xscans}
\end{figure}

For each scan head, several scan routes are conducted within its subspace. A scan route records an actual traversal path of the image patches. In the same scan pattern, we can start scanning from patches located at different positions (such as corners), so there will be multiple scan routes. We use four routes in a scan head by default. An example is shown in \figref{fig:xscans}. These scan routes are used to gather embedding sequences which are transformed through the Mamba blocks. For instance, four scan routes produce four embedding sequences. Each sequence is reorganized as an embedding section by stacking the embedding vectors along the predefined traversal path. Thereafter ESF is applied to fuse these embedding sections to produce the final patch embeddings. In addition to direct addition $z_1 = \sum_{i=1}^{k} y_i$, we further introduce two alternative schemes.

\begin{figure}[!ht]
\centering
\subfigure[]{
\begin{minipage}[b]{0.723\linewidth}
\includegraphics[width=\linewidth]{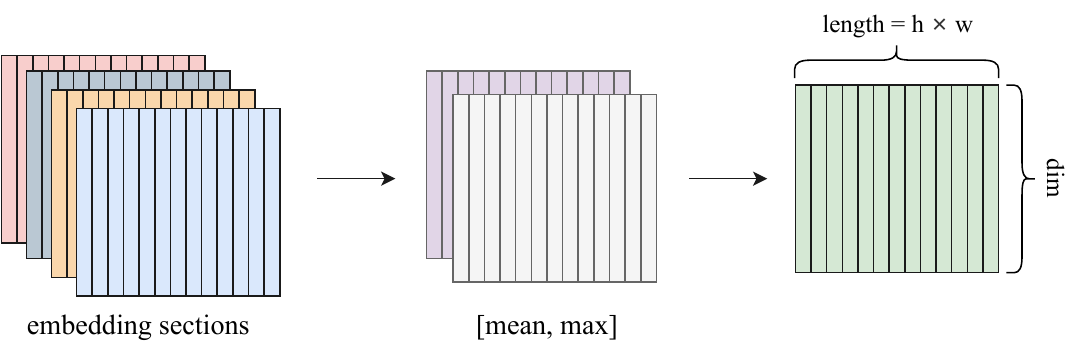} 
\end{minipage}
}
\hspace{0.05in}
\subfigure[]{
\begin{minipage}[b]{0.235\linewidth}
\includegraphics[width=\linewidth]{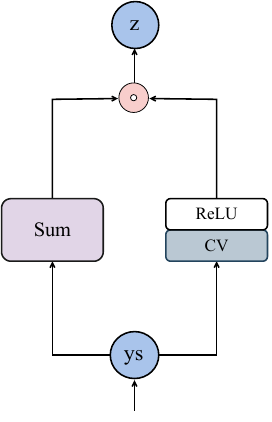}
\end{minipage}
}  
\caption{
Illustrations of two schemes for the ESF sub-module. (a) Mixture of Poolings; (b) CV-guided Scaling.
}
\label{fig:esfsubmodule}
\end{figure}

\textbf{Mixture of Poolings.} Here we partially draw on the design of spatial attention module in CBAM \cite{woo2018cbam}. As shown in \figref{fig:esfsubmodule}(a), we first compute the mean value and maximum value along the sectional axis, thus obtain two pooling sections, and concatenate them to generate a concise feature descriptor, then forward them to a linear layer to efficiently output the final embedding section. This process can be formulated as:

\begin{equation}
z_2 = W[x_0, x_1]^{T}
\end{equation}

where $W \in R^{1 \times 2}$, $x_0 = \frac{1}{k}\sum_{i=1}^{k} y_i $ and $x_1 = max([y_0,y_1,\cdots,y_k])$ are the average pooling and max pooling along the section axis respectively. On the concatenated feature descriptor $[x_0, x_1]$, we apply a linear projection to generate the fused embedding section.

\textbf{CV-guided Scaling.} In Transformers, positional encoding plays a crucial role in providing the model with information about the relative or absolute position of the tokens in the sequence. Instead of using explicit positional encoding, we introduce a mechanism to enhance the implicit perception ability regarding the spatial positions of image patches. Specifically, to allow the model screen out trivial features which are insensitive to the scan routes, we introduce a scan route attention mechanism. For each scan pattern, we select $k$ scan routes which start scanning from different corner patches, thus we obtain $k$ scan sequences. There are $k$ values for each component of the embedding of each patch. To quantify the awareness of extracted embeddings to the scan route, we calculate a variant of the Coefficient of Variation (CV) for these $k$ values. Coefficient of variation is a measure of relative variability or dispersion of data around the mean value in a sample or population. If these $k$ values from $k$ scan routes are more consistent, thus the CV is relatively small. On the contrary, if these values are more dispersed, then the CV is relatively large. To a certain extent, the embedding extracted by these scan routes is well aware of positional information and is worth preserved or enhancing. We formulate the above process as:

\begin{equation}
z_3 = (\sum_{i=1}^{k} y_i) \odot \sigma(y_{cv})
\end{equation}

where $y_{cv} = \operatorname{std}([y_i]) / \operatorname{avg}([y_i-\operatorname{min}([y_i])])$ is a variant of the CV, and $\odot$ denotes the element-wise product between tensors, and $\sigma(x)$ is a monotone function, such as Sigmoid and ReLU. This monotone function is introduced to prompt the Mamba block to extract position-aware features. A simplified flowchart is depicted in \figref{fig:esfsubmodule}(b).

In our actual experiment, a parameter $t$ is introduced to filter out features with lower CV values.

\begin{equation}\label{eqn:cvscaling}
\sigma(x, t) = \operatorname{ReLU}(x-t) = \operatorname{max}(0, x-t)
\end{equation}

This function returns $0$ when $x < t$ and $x-t$ otherwise. The parameter $t$ can be set as a hyperparameter or a learnable parameter. Such a strategy can be considered as a novel regularization technique to prevent over-fitting and improve generalization. 

Furthermore, it is also feasible to merge last two schemes, resulting in a more complicated scheme:

\begin{equation}
z_4 = z_2 \odot \sigma(y_{cv})
\end{equation}

However, the actual experiments on the ISIC18 dataset reveal no significant performance improvement, while they demonstrate an increase in computational overhead.

Note that the above-mentioned aggregation operations (such as average, min, max and std) are performed along the section axis.

\subsection{Projection}

Finally, the tail part is designed to transform the sub-embeddings back into high dimensional spaces. To this end, we first concatenate the sub-embeddings, then normalize them by Layer Normalization, and finally project the normalized results back into the high dimensional space. In practice, if the sum of the dimensions of these subspace $n \times S$ is equal to the one of input embedding space $C_1$, the last projection can optionally be excluded from the architecture. The performance comparison is illuminated through the ablation study in the following section.

We have also substituted the projection component with a feed-forward network (ffn); however, experiments conducted on small datasets indicate that this modification increases the parameter count without substantially enhancing segmentation performance.

\subsection{Pseudo-code for MHS Module}

Given an input image $I \in R^{H \times W \times D}$, where $H$, $W$, and $D$ denote height, width and the number of channels of the image, respectively. First, a Patch Embedding layer is applied to divide the input image into non-overlapping patches of size $4 \times 4$, subsequently our MHS modules are alternated with patch merging layers to transform them into a series of embedding spaces $R^{\frac{H}{4} \times \frac{W}{4} \times C}$, $R^{\frac{H}{8} \times \frac{W}{8} \times 2C}$, $R^{\frac{H}{16} \times \frac{W}{16} \times 4C}$, with $C$ defaulting to $96$ as in VM-UNet. One of our modules takes a $C_l$-channel embedding map as input and produces an output embedding map with an equivalent number of channels. Specifically, the pseudo-code for the proposed MHS module is presented in \textbf{Algorithm 1}. 

\begin{algorithm}[h]
\caption{Pseudo-code for MHS module}
\small
\renewcommand{\algorithmicensure}{\textbf{Output:}}
\begin{algorithmic}[1]
\renewcommand{\algorithmicrequire}{\textbf{Input:}}
\REQUIRE {embeddng map $\mathbf{X}_{l-1}$: \textcolor{shapecolor}{$(\mathtt{B}, \mathtt{H}, \mathtt{W}, \mathtt{C})$}}
\ENSURE {embedding map $\mathbf{X}_{l}$ : \textcolor{shapecolor}{$(\mathtt{B}, \mathtt{H}, \mathtt{W}, \mathtt{C})$}}
\STATE \textcolor{lgreen}{\text{/* reshape the input map */}}
\STATE $\mathbf{x}$ : \textcolor{shapecolor}{$(\mathtt{B}, \mathtt{C}, \mathtt{H}, \mathtt{W})$} $\leftarrow$ $\mathbf{Reshape}(\mathbf{X}_{l-1})$
\STATE \textcolor{lgreen}{\text{/* process with $n$ Scan Heads */}}
\FOR{$h$ in \{$1$, $2$, $\cdots$, $n$\}}
    \STATE \textcolor{lgreen}{\text{/* project onto a subspace */}}
    \STATE $\mathbf{x}^h$ : \textcolor{shapecolor}{$(\mathtt{B}, \mathtt{S}, \mathtt{H}, \mathtt{W})$} $\leftarrow$ $\mathbf{Linear}_h(\mathbf{x})$
    \STATE \textcolor{lgreen}{\text{/* $K$ Scan Routes of the $h$-th Scan Head */}}
    \FOR{$j$ in \{$1$, $2$, $\cdots$, $K$\}}
        \STATE $\mathbf{x}^h_j \;$ : \textcolor{shapecolor}{$(\mathtt{B}, \mathtt{S}, \mathtt{HW})$} $\leftarrow$ $\mathbf{Scan\_Route}_j(\mathbf{x}^h)$ 
        \STATE ${\mathbf{x}'}^h_j$ : \textcolor{shapecolor}{$(\mathtt{B}, \mathtt{S}, \mathtt{HW})$} $\leftarrow$ $\mathbf{Mamba}_{h} (\mathbf{x}^h_j)$
        \STATE ${\mathbf{y}}^h_j \;$ : \textcolor{shapecolor}{$(\mathtt{B}, \mathtt{S}, \mathtt{HW})$} $\leftarrow$ $\mathbf{Rearrange} ({\mathbf{x}'}^h_j)$
    \ENDFOR
    \STATE $\mathbf{Y}^h$ : \textcolor{shapecolor}{$(\mathtt{B}, \mathtt{K}, \mathtt{S}, \mathtt{HW})$} $\leftarrow$ $\mathbf{Cat}([\mathbf{y}^h_1,\mathbf{y}^h_2,\cdots,\mathbf{y}^h_K)]$
    \STATE \textcolor{lgreen}{\text{/* Embedding Section Fusion */}}
    \STATE $\mathbf{y}^h_{cv}$ : \textcolor{shapecolor}{$(\mathtt{B}, \mathtt{S}, \mathtt{HW})$} $\leftarrow$ $\mathbf{Coefficient\_Variation}(\mathbf{Y}^h)$ 
    \STATE $z^h$ : \textcolor{shapecolor}{$(\mathtt{B}, \mathtt{S}, \mathtt{HW})$} $\leftarrow$ 
 $\mathbf{Sum}(\mathbf{Y}^h) \bigodot \;\! \mathbf{ReLU}(\mathbf{y}^h_{cv}-t)$ 
\ENDFOR
\STATE \textcolor{lgreen}{\text{/* concatenate along the channel axis */}}
\STATE $\mathbf{y}$ : \textcolor{shapecolor}{$(\mathtt{B}, \mathtt{nS}, \mathtt{HW})$} $\leftarrow$ $\mathbf{Layer\_Norm}(\mathbf{Cat}([\mathbf{z}^1,\mathbf{z}^2,\cdots,\mathbf{z}^n]))$
\STATE \textcolor{lgreen}{\text{/* projection, this step is optional */}}
\STATE $\mathbf{y}$ : \textcolor{shapecolor}{$(\mathtt{B}, \mathtt{C}, \mathtt{HW})$} $\leftarrow$ $\mathbf{Linear}(\mathbf{y})$
\STATE \textcolor{lgreen}{\text{/* reshape the output embedding map */}}
\STATE $\mathbf{X}_{l}$ : \textcolor{shapecolor}{$(\mathtt{B}, \mathtt{H}, \mathtt{W}, \mathtt{C})$} $\leftarrow$ $\mathbf{Reshape}(\mathbf{y})$
\STATE \textcolor{lgreen}{\text{/* return $\mathbf{X}_{l}$ */}}
\STATE Return: $\mathbf{X}_{l}$ 
\label{alg:block}
\end{algorithmic}
\end{algorithm}

\section{Experimental results}

In this section, we conduct experiments of our module for medical image segmentation tasks, especially on relatively small datasets. To assess the performance of our module, we apply it to VM-UNet, a recently proposed Mamba-based framework for medical image segmentation. 

\textbf{It is crucial to emphasize that we do not incorporate additional mechanisms to enhance the model's performance, as in the case of VM-UNet v2 \cite{zhang2024vmunetv2} and UltraLight VM-UNet \cite{wu2024ultralight}. Our efforts are centered on harnessing the potential of the vision Mamba further. To verify the efficacy of our module, we preserve the rest of network architecture of VM-UNet unchanged, and we just replace the SS2D module in VSS block with our MHS module (see \figref{fig:modules}).}

\subsection{Datasets and Experimental Setups}
Specifically, we evaluate the performance of our module on three publicly available medical image datasets: ISIC17 \cite{isic17}, ISIC18 \cite{isic18}, and Synapse \cite{synapse2015} datasets. These datasets, which are widely used in current segmentation research, are employed in this study to benchmark the competitive performance of the proposed module.

We implemented MHS-VM using the PyTorch 2.0 and trained the networks with the same parameters as VM-UNet. Batch size is set to 32, and training epochs are set to 300. The AdamW \cite{adamw} optimizer is employed with an initial learning rate of 1e-3. CosineAnnealingLR \cite{cosineannealingLR} is utilized as the scheduler with a maximum of 50 iterations and a minimum learning rate of 1e-5. All experiments are conducted on a single NVIDIA RTX 4090 GPU.

\subsection{Ablation Study}

In this section, we conduct a series of ablation experiments to explored various configurations for the proposed MHS module using the ISIC18 dataset. To assess the performance of our module independently, we train all our networks from scratch, as opposed to employing any pre-trained weights.

\textbf{Effect of ESF.} This sub-module is specifically crafted to integrate the embedding sections extracted from the diverse scan routes within a single scan head. We evaluated the three aforementioned schemes and the results revealed that the third scheme outperformed the others, achieving the highest performance. The experimental results are presented in \tabref{tab:esfablation}. In this experiment, we deployed three scan heads that correspond to the last three scan patterns depicted in \figref{fig:scanpatterns}, and designated $t=0.5$ in Eqn. \eqref{eqn:cvscaling} as a hyperparameter. In subsequent experiments, we will primarily employ the third scheme (CV-guided scaling) in ESF sub-modules.

\begin{table}[!h]
    \centering
    \setlength\tabcolsep{3pt}
	\renewcommand\arraystretch{1.25}
 	\caption{Ablation studies on various ESF schemes.}
\begin{tabular}{cccc|cccc|cccc}
\hline
 \multicolumn{4}{c|}{Sum($+$)} & \multicolumn{4}{c|}{Mixture of Poolings} & \multicolumn{4}{c}{CV-guided Scaling} \\ 
             mIoU(\%) & DSC(\%) & Params & FLOPs  & mIoU(\%)       & DSC(\%) & Params & FLOPs & mIoU(\%) & DSC(\%) & Params & FLOPs      \\ \hline
79.45 & 88.55 &17.3975&2.3902&79.64 &88.65 &17.3976&2.3953& 79.91 &88.83 &17.3975 &2.3902 \\
\hline
\end{tabular}
    \label{tab:esfablation}
\end{table}

\textbf{Effect of projection.} We further remove the projection in tail part of our MHS module. The removal of this component results in the substantial reduction in the parameter count and computational overhead, but the performance remains comparable (see \tabref{tab:projection}). Consistent with the previous setup, we continue to employ three scan heads for this comparative experiment.

\begin{table}[!h]
    \centering
    \setlength\tabcolsep{3pt}
	\renewcommand\arraystretch{1.25}
 	\caption{Ablation studies on Projection. (with \raisebox{-0.5ex}{\CheckmarkBold} or without \raisebox{-0.5ex}{\XSolidBrush})}
\begin{tabular}{c|cccc|cccc}
\hline
\multirow{2}{*}{Dataset} & \multicolumn{4}{c|}{Projection \raisebox{-0.5ex}{\CheckmarkBold}} & \multicolumn{4}{c}{Projection \raisebox{-0.5ex}{\XSolidBrush}} \\ 
             & mIoU(\%) & DSC(\%) & Params(M) & FLOPs(G)    & mIoU(\%) & DSC(\%) & Params(M) & FLOPs(G)         \\ \hline
ISIC18 &79.91 &88.83 & 17.3975 & 2.3902 &79.86 &88.80 & 14.2619 & 1.8137 \\
\hline
\end{tabular}
    \label{tab:projection}
\end{table}

The projection in the tail part is employed to amalgamate the components of sub-embeddings. Given that the head in the subsequent layer also incorporates the projections, we can eliminate the projection at the tail of the current layer, thereby substantially reducing the parameter count. However, if the sum of the dimensions of the subspaces does not match the dimension of the embedding space, the projection cannot be omitted arbitrarily.

\textbf{Number of Scan Heads.} In the previous experiments, we focus on experimenting with three scan heads to gather features in a 2D image. Next, we investigate whether more scan heads enhances the performance of our model. Empirical evaluations reveal that while increasing the number of scan heads can marginally enhance model performance, the improvement is quite modest. Nonetheless, the number of parameters has slightly decreased.

Furthermore, we conduct experiments to explore the effects of increasing the dimensionality of the parallel subspace. With the projection, the dimension of subspace is unrestricted, thus allowing for the possibility of further exploration of its potential. As shown in \tabref{tab:heads}, experiments on the ISIC18 dataset indicate that the performance of the model achieves a slight enhancement, accompanied by a significant increase in the parameter count. Augmenting the dimension of the subspace may yield advantages for datasets with a greater volume of data, an area that warrants further investigation in the future.

\begin{table}[!h]
    \centering
    \setlength\tabcolsep{3pt}
	\renewcommand\arraystretch{1.25}
 	\caption{Ablation studies on different number of scan heads.}
\begin{tabular}{c|cc|cc}
\hline
\multirow{2}{*}{Num of Heads} & \multicolumn{2}{c|}{CV-guided Scaling} & \multirow{2}{*}{Params(M)} & \multirow{2}{*}{FLOPs(G)} \\ 
             & mIoU(\%)        & DSC(\%)        \\ 
             \hline
$n = 3, S=C_l/3$ & 79.91 & 88.83 &17.3975 &2.3902\\
\hline
$n = 4, S=C_l/4$ & 79.96 & 88.86 &17.2685 &2.3902 \\
\hline
$n = 4, S=C_l/3$ & 80.16 & 88.99 &20.0352 &2.5823 \\
\hline
\end{tabular}
    \label{tab:heads}
\end{table}
 
For reference, the original VM-Unet comprises 27.4276M parameters with 4.1119G FLOPs. In the case of employing four heads, we utilized a configuration that encompasses all scan patterns depicted in \figref{fig:scanpatterns}, with each pattern being associated with one of the four scan heads. Consistent with the previous experiments, we still set $t=0.5$ in Eqn. \eqref{eqn:cvscaling}. In addition, we also used the second model with the fewest parameters in the table to experiment on the ISIC17 dataset and obtained scores of 78.97\% (mIoU) and 88.25\% (DSC). For this dataset, the model performs well when $t$ is set to $0$. To optimize performance across different datasets, the value of the hyperparameter $t$ may require calibration.

\subsection{Comparative Results}

We employ the most lightweight model presented in \tabref{tab:projection} for comparison with the original VM-UNet. The comparative experimental results presented in \tabref{tab:comparison} underscore that our VM-UNet variant attains superior performance on these datasets. It is crucial to highlight that we exclusively substituted the SS2D block in VM-UNet with our proposed module, thereby isolating the observed improvements in model performance as a clear indication of the efficacy of our module. 

\begin{table}[!h]
    \centering
    \setlength\tabcolsep{3pt}
	\renewcommand\arraystretch{1.25}
 	\caption{Comparative experimental results on three datasets using various models.}
\begin{tabular}{c|cc|cc|cc|cc}
\hline
\multirow{2}{*}{Models} & \multirow{2}{*}{Params(M)}&
\multirow{2}{*}{FLOPs(G)} & \multicolumn{2}{c|}{ISIC17} & \multicolumn{2}{c|}{ISIC18} & \multicolumn{2}{c}{Synapse} \\ 
& & &
             mIoU(\%)$\uparrow$     &
             DSC(\%)$\uparrow$      &
             mIoU(\%)$\uparrow$     &
             DSC(\%)$\uparrow$      &
             mIoU(\%)$\uparrow$     & 
             HD95$\downarrow$      \\ 
             \hline
VM-UNet & 27.4276 & 4.1119 &77.59 &87.38 &78.66 &88.06 & 75.50 & 36.41 \\
VM-UNet-T & 27.4276 & 4.1119 & 78.85 & 88.17 & 79.04 & 88.29 & - & - \\ 
MHS-UNet & 14.2619 & 1.8137 & 78.81 & 88.15 & 79.85 & 88.67 & 77.13 & 27.74 \\
\hline
\end{tabular}
    \label{tab:comparison}
\end{table}

The VM-UNet-T model is trained by initializing VM-UNet with the pre-trained weights obtained from VMamba-T \cite{liu2024vmamba}. The VM-UNet model and our model (MHS-UNet for short) are trained from scratch for 300 epochs. Significantly, the updated network, compared to the original VM-UNet, not only achieves superior performance but also exhibits a remarkable reduction in parameter count and computational overhead by 48.00\% and 55.89\%, respectively. The performance of our model either approaches or even surpasses that of VM-UNet-T initialized with pre-trained weights. However, there is a certain gap from the performance of VM-UNet-S which is trained by initializing VM-UNet with the pre-trained weights obtained from VMamba-S \cite{liu2024vmamba}.

\section{Conclusion and Future Work}

To enhance the performance of Mamba for visual tasks, we introduce a multi-head scan module based on Vision Mamba, abbreviated as MHS-VM. We project the patch embedding into several parallel subspaces and introduce various scan patterns to capture complex dependencies of patches in a 2D image. To verify the efficacy of our module, we replace the SS2D module in VM-UNet with our module, keep the rest of the network architecture unchanged, and then conduct ablation studies based on this framework. Relative to the original VM-UNet, the revised network incorporating our module features a reduced parameter count and computational overhead, and our experiments conducted on three publicly available datasets confirm that our module enhances prediction performance.

While the use of 1D selective scan for 2D visual tasks emerges as a promising approach meriting further exploration, it is important to note that the patterns discussed in this paper are not exhaustive. The potential for discovering and implementing additional scan patterns offers a rich terrain for future research, opening up avenues for even more sophisticated and nuanced applications in the field of visual recognition and processing. We will ameliorate the architecture and parallel programming, and further explore hierarchical representations that integrate local scanning and global scanning, thereby make our module a general-purpose backbone for more visual tasks in the future.



\bibliographystyle{plain}
\bibliography{related}  

\end{document}